\documentclass[%
 reprint,
 amsmath,amssymb,
 aps,
]{revtex4-2}

\usepackage{graphicx}
\usepackage{dcolumn}
\usepackage{bm}
\usepackage{subcaption}

\begin{document}

\preprint{APS/123-QED}

\title{Fast Posterior Probability Sampling with Normalizing Flows and Its Applicability in Bayesian analysis in Particle Physics}

\author{Mathias El Baz}
 \email{mathias.elbaz@unige.ch}

\author{Federico Sánchez}
\affiliation{
Département de Physique Nucléaire et Corpusculaire, Université de Genève}%

\date{\today}

\begin{abstract}
In this study, we use Rational-Quadratic Neural Spline Flows, a sophisticated parametrization of Normalizing Flows, for inferring posterior probability distributions in scenarios where direct evaluation of the likelihood is challenging at inference time. We exemplify this approach using the T2K near detector as a working example, focusing on learning the posterior probability distribution of neutrino flux binned in neutrino energy. The predictions of the trained model are conditioned at inference time by the momentum and angle of the outgoing muons released after neutrino-nuclei interaction. This conditioning allows for the generation of personalized posterior distributions, tailored to the muon observables, all without necessitating a full retraining of the model for each new dataset. The performances of the model are studied for different shapes of the posterior distributions. 

\end{abstract}

\maketitle


\section*{\label{sec:level1}Introduction}
Particle physics experiments often face the challenge of deducing hidden properties underlying the collected data. For instance, when studying neutrinos, statistical methods are employed to estimate latent variables like neutrino masses and mixing angles. 

In this context, both frequentist and Bayesian methods are used and even sometimes cross-validated. The frequentist approach focuses on deriving estimates by analyzing the observed data, assuming that the latent variables have fixed, but unknown, values.  This implies searching for a single set of latent variable values that is inferred with tools such as maximum likelihood estimation. On the other hand, the Bayesian approach does not assume fixed latent variables, but aims to estimate their posterior distribution. This is particularly advantageous when dealing with complex parameter spaces, common in particle physics, as it facilitates the quantification of uncertainty and the exploration of parameter correlations. 

However, the implementation of Bayesian methods is not without challenges, especially in situations where sampling the posterior probability distribution is required to be fast. Traditional Bayesian methods such as Variational Inference (VI) or Markov Chain Monte-Carlo (MCMC) rely heavily on the likelihood estimation, often rendering them ineffective when the density estimation is slow. Moreover, these methods suffer from long ”burn-in” or ”training” times, making the overall sampling process very slow. 

The rapid advancements in Artificial Intelligence, particularly in Deep Learning have unlocked new avenues for fast and flexible Bayesian inference. In this context, our paper explores an alternative machine-learning-based method for Bayesian inference, using a model based on Conditional Normalizing Flows (CNF). Normalizing flows- the foundation of CNF- model complex distributions by transforming a simple distribution like a normal distribution into a more intricate one. This is achieved through a series of learnable transformations that progressively shape the simple distribution into one that closely resembles the target distribution. CNF extends this concept by conditioning the flow transformation on the given dataset of observed variables at inference time. This adaptability allows for the elimination of the need for retraining with new datasets, making the sampling process more effective. Furthermore, unlike traditional likelihood-based Bayesian methods like MCMC, CNF does not require the evaluation of the likelihood at inference time, thereby making the sampling faster when the calculation of the likelihood is time consuming. 

To illustrate this method, we will use the near detector fit of the T2K (Tokai to Kamioka)  experiment \cite{T2K:2011qtm} as a working example for our CNF model. At the near detector, muon neutrinos interact with nuclei releasing muons observed further. In our exploratory approach, we aim at estimating the posterior distributions of the neutrino energy bin values (the latent variables) from a dataset of muon momenta and angles (the observed variables). The variance of the posterior probability density translates the uncertainty created by the Poissonian statistics when inferring the latent variables. A detailed problem description is given in Section \ref{problem definition}. Section \ref{Conditional density estimation with Normalizing Flows} introduces the main concepts of conditional probability estimation using normalizing flows. We apply this concept to a simplified version of the near detector fit in Section \ref{Amortized Normalizing Flows for the Near Detector fit at T2K}. Finally, Section \ref{Quantifying and Modeling Non-Gaussian Characteristics} is an exploration of how CNFs behave when tasked to predict a more complex posterior distribution.

\section{Problem definition}
\label{problem definition}

The T2K experiment investigates neutrino oscillations. The principal challenge encountered by the T2K collaboration is the complex parametrization of the models of neutrino cross-section, neutrino flux as a function of the neutrino energy and detector response used to infer the oscillation parameters. One objective of the near detector data fit is to constrain the neutrino cross-section and flux models. The near detector data fit involves searching for an optimal set of parameters describing how systematic uncertainties change the predictions on the event rate, given the data. So far, only the outgoing muon observables $(p_\mu,\theta_\mu)$ after interaction between the neutrino and the nuclei are used for the near detector fit at T2K. The near detector data fit is achieved by studying a binned Barlow-Beeston\cite{Barlow:1993dm} likelihood with penalty terms corresponding to the Gaussian priors of the systematics parameters of the flux, cross-section, and detector response models \cite{T2K:2023smv} jointly noted  $\vec{s}$:
\begin{eqnarray*}
-\ln (L)  & = \sum\limits_i \beta_i N_i^p(\vec{s})-N_i^d+N_i^d \ln \left(\frac{N_i^d}{\beta_i N_i^p(\vec{s})}\right) + \frac{(\beta_i-1)^2}{2\sigma_{\beta_i}^2}  \notag \\ & +\sum\limits_i^{\vec{s}} \sum\limits_j^{\vec{s}} \Delta(\vec{s})_i\left(V_{\vec{s}}^{-1}\right)_{i, j} \Delta(\vec{s})_j
\label{likelihood equation}
\end{eqnarray*}
where the sum is over the energy bins, $N_i^{\text{obs}}$ is the observed number of events and $N_i^{\text{pred}}$ is the predicted number of events in the $i$-th bin. The Barlow-Beeston likelihood introduces parameters $\beta_i$ that account for the Poisson fluctuations in the Monte-Carlo dataset that yields to $N_i^{\text{pred}}$. The second line captures the Gaussian uncertainties arising from the systematics $\vec{s}$ with covariance matrix $V_{\vec{s}}$.

T2K's near detector fit uses a frequentist method called BANFF \cite{Walsh:2022abr} and a Bayesian method rooted in MCMC called Mach3 \cite{Sztuc:2020qff}. In our exploratory method, we exclude the T2K systematics that account for cross-sections, flux and detector uncertainties. Furthermore, the remaining latent variable, represented by the vector of Barlow-Beeston reweights $\beta$, only conditions the marginal distribution of \(E_\nu\). Therefore, the joint probability distribution of seeing an event with $(p_\mu,\theta_\mu)$ is as follows:

\[p(p_\mu, \theta_\mu|\mathbf{\beta}) = \int p(p_\mu, \theta_\mu| E_\nu) \times p(E_\nu|\mathbf{\beta}) dE_\nu\]

We can note that the likelihood $L$ is formulated with a straightforward linear dependence on the Barlow-Beeston parameters. Nevertheless, the versatility of neural networks enables them to capture more complex relationships between latent variables and observations. This capability suggests the potential applicability of our approach to more complex systematic uncertainties, such as energy shifts \cite{Chakrani:2023htw}. One can envision a scenario where we simply rewrite the expression of the likelihood function, incorporating a more intricate formulation of the $p(E_\nu, p_\mu, \theta_\mu|\mathbf{\beta})$ distribution. However, delving into these more elaborate systematics extends beyond the scope of the current study.

Given a dataset denoted as $\mathbf{X}_d$, consisting of events sampled from the distribution $p(p_\mu, \theta_\mu|\mathbf{\beta})$, the objective of our deep learning model is to predict the corresponding posterior distribution $p(\beta|\mathbf{X}_d)$ during the inference phase. The vector $\beta$, representing Barlow-Beeston reweights, is the only latent variable in this context. Hence, the spread of the posterior distribution is solely a reflection of the Poisson statistics. Our goal is to design a model to perform Bayesian inference automatically at inference time for a diverse range of datasets $\mathbf{X}_d$.

As Papamakarios outlines in his thesis \cite{papamakarios2019neural}, Bayesian techniques can be divided into two categories: "likelihood-based inference" methods, which require an estimation of the likelihood density $p(\mathbf{X}_d|\beta)$ during inference, and "likelihood-free inference" methods, which do not.

By binning the $(p_\mu, \theta_\mu)$ space, we can estimate the Poisson negative log-likelihood given the observation of a dataset $\mathbf{X}_d$ as follows:

\begin{equation}
    p(\mathbf{X}_d|\beta) = \sum_{(i,j)} \left[ N_{(i,j)}^p(\beta) - N_{(i,j)}^d + N_{(i,j)}^d \ln \left( \frac{N_{(i,j)}^d}{N_{(i,j)}^p(\beta)} \right) \right]
    \label{Poisson likelihood}
\end{equation}

Here, the summation is over bins $(i,j)$, and $N_{(i,j)}^d$ and $N_{(i,j)}^p(\beta)$ denotes respectively the detected and predicted number of events in bin $(i,j)$ for the reweight $\beta$. Therefore, using the likelihood to train and test our Deep Learning model is feasible. However, this calculation is computationally intensive, especially for a significant number of bins or in a high-dimensional latent space. Therefore, we opted for an alternative method that does not require likelihood estimation during training and inference. This method will be further discussed in the following sections and in Appendix \ref{Generating the datasets}.

\section{Conditional density estimation with Normalizing Flows}
\label{Conditional density estimation with Normalizing Flows}

 Normalizing Flows (NF) are often used to estimate a target distribution. In this particular context, the adoption of NF emerges organically as a means to predict a posterior distribution $q_\phi(\beta | \mathbf{X}_d)$ close to the true posterior distribution $p(\beta | \mathbf{X}_d)$, where $\phi$ are the parameters of the NF model. The basic concepts of Normalizing Flows will be explained in Section \ref{Normalizing Flows}. A more complete overview of Normalizing Flows can be found in the review of Papamakarios et al \cite{papamakarios2021normalizing}. In this work, the density estimation is performed through Rational-Quadratic Neural Spline Flows, a specific implementation of Normalizing Flows that will be presented in Section \ref{Neural Spline Flows}. Finally, Section \ref{Normalizing Flows for Amortized Variational Inference} describes the utilization of Normalizing Flows for Bayesian inference, explaining the concept of "Conditional Normalizing Flows" as a means to learn a posterior distribution without the need of re-training the model for new datasets $\mathbf{X}_d$.

\subsection{Normalizing Flows}
\label{Normalizing Flows}

\subsubsection{Definition}

At its core, a Normalizing Flow is a diffeomorphism of the probabilistic space that transforms a simple probability distribution into a more complex one. The concept revolves around the simple change of variable rule in probability theory. 

\noindent For a diffeomorphism of random variables, from \( \mathbf{x} \) to \( \mathbf{z} \), where \( \mathbf{z} = T(\mathbf{x})\), the probability density function of \( \mathbf{z} \) is related to the one of \( \mathbf{x} \) as follows:
\[ p_{\mathbf{z}}(\mathbf{z}) = p_{\mathbf{x}}(\mathbf{x}) \left| \det \left( \mathbf{J}_T(\mathbf{x}) \right) \right|^{-1} \]
with  $J_T(\mathbf{x})$ the Jacobian of the transformation. $p_{\mathbf{x}}$ represents here the base distribution (a normal distribution in general) from where we model the target distribution $p_{\mathbf{z}}$ by applying the diffeomorphism $T$ to the probabilistic place. 

The flow transformation can be composed of multiple NFs. Suppose we represent the transformation \( T \) as a composition of simpler \( T_k \) transformations, with \( T = T_K \circ \cdots \circ T_1 \). Starting with an initial value \( \mathbf{z}_0 = \mathbf{x} \) and target value \( \mathbf{z}_K = \mathbf{z} \), we can evaluate the transformation and compute the Jacobian as follows:
\[
\begin{aligned}
\mathbf{z}_k  = T_k&\left(\mathbf{z}_{k-1}\right), \quad k = 1: K, \\
\left|J_T(\mathbf{z})\right| & = \left|\prod_{k=1}^K J_{T_k}\left(\mathbf{z}_{k-1}\right)\right|,
\end{aligned}
\]
where \( J_{T_k}\left(\mathbf{z}_{k-1}\right) \) represents the Jacobian determinant of the \( T_k \) transformation evaluated at \( \mathbf{z}_{k-1} \). In practical applications of Normalizing Flows, the transformations \( T_k \) (or \( T_k^{-1} \)) are often implemented using a neural network, which provides the required flexibility to model complex mappings. 

\subsubsection{Loss function}

The training of neural networks requires a loss function to estimate the divergence between the predicted and true distributions.  The Kullback-Leibler (KL) divergence is a fundamental concept in statistics, widely used to compare two probability distributions \cite{KL_1951}. The key intuition behind using KL-divergence lies in its ability to quantify the information lost when one distribution is used to approximate another.  Formally, for two probability distributions P and Q over the same space $\Omega$, the KL-divergence is defined as :
\begin{equation}
    D_{\text{KL}}(P || Q) = \sum_{x \in \Omega} P(x) \log \left( \frac{P(x)}{Q(x)} \right)
    \label{KL}
\end{equation}

An essential characteristic of the KL-divergence is that it is non-negative, with \(D_{\text{KL}}(P || Q) = 0\) if and only if \(P\) and \(Q\) are identical distributions. This property allows us to derive a loss function from the KL-divergence as:
$$
\begin{aligned}
L(\phi) & =D_{\mathrm{KL}}\left(p_\mathbf{z}(\mathbf{z}) \| q_\phi(\mathbf{z})\right) \\
& = \mathbb{E}_{\mathbf{z} \sim p_\mathbf{z}(\mathbf{z})} [\log p_\mathbf{z}(\mathbf{z}) -\log  q_\phi(\mathbf{z})]\\
& = \mathbb{E}_{\mathbf{z} \sim p_\mathbf{z}(\mathbf{z})} [\log p_\mathbf{z}(\mathbf{z}) ] \\
& \quad -\mathbb{E}_{\mathbf{z} \sim p_\mathbf{z}(\mathbf{z})}\left[\log p_{\mathbf{x}}\left(T^{-1}(\mathbf{z} ; \phi)\right)+\log \left|\operatorname{det} J_T^{-1}(\mathbf{z} ; \phi)\right|\right]
\end{aligned}
$$
where $\mathbb{E}_{\mathbf{z} \sim p_\mathbf{z}(\mathbf{z})}$ is the expectation for samples of the target distribution $p_\mathbf{z}$, $\phi$ represents the parameters of the flow $T$ parametrized by the neural network, $p_{\mathbf{x}}$ is the base distribution and $q_\phi$ is the predicted distribution. 

The loss function can be computed for target densities \(p_\mathbf{z}\) from which one can sample, but the density evaluation for a specific point \(z\) is not required. When optimizing the transformation \(T\), we estimate the gradient of the KL-divergence by drawing samples from the target distribution ${\mathbf{z}_n} \sim p_\mathbf{z}(\mathbf{z})$ :
\begin{eqnarray}
\nabla_\phi L(\phi) & \approx  -\frac{1}{N} \sum_{n=1}^N[\nabla_\phi \log p_{\mathbf{x}}\left(T^{-1}\left(\mathbf{z}_n ; \phi\right)\right)\notag\\
& \;+\nabla_\phi \log \left|\operatorname{det} J_T^{-1}\left(\mathbf{z}_n; \phi\right)\right|]
\label{Dkl gradient}
\end{eqnarray}
\subsubsection{Autoregressive flows}

For a transformation to be valid, it must be both invertible and possess a tractable Jacobian determinant. However, even if a network ensures theoretical invertibility, practical computations might still be expensive or infeasible. The computation of the determinant of the Jacobian is required for the loss calculation and sampling scales with a cubic complexity of the probability space dimension. A tractable Jacobian determinant implies a complexity that scales at most $\mathcal{O}(D)$, where $D$ is the dimension of the probabilistic space, which is ensured by autoregressive flows \cite{kingma2017improving}. 

In autoregressive flows, the transformation is structured such that each output dimension depends only on its lower dimensions. This involves employing D component-wise transformations, referred to as transformers, such as:
\[
z_i^{\prime}=\tau\left(z_i ; \mathbf{h}_i\right) \text{ with } \mathbf{h}_i=\mathbf{h}_i\left(\mathbf{z}_{<i} ; \phi\right)
\]
where \(z_i^{\prime}\) is the \(i\)-th component of \(\mathbf{z}^{\prime}\), \(z_i\) is the \(i\)-th component of \(\mathbf{z}\), \(\tau\) represents the transformer, which is a one-dimensional diffeomorphism concerning \(z_i\) and depends on the \(i\)-th dimension of the conditioner \(\mathbf{h}_i\).  \(\mathbf{h}_i\) takes \(\mathbf{z}_{<i}=\left(z_1, z_2, \ldots, z_{i-1}\right)\) as input, i.e., the previous components of \(\mathbf{z}\).

Due to this definition, the Jacobian matrix, denoted as \(J_{T}(\mathbf{z})\), is lower triangular. Consequently, the log-determinant of the Jacobian matrix can be efficiently computed by taking the sum of the logarithm of its diagonal elements:
\[
\log \left|\operatorname{det} J_{T}(\mathbf{z})\right|=\sum_{i=1}^D \log \left|\frac{\partial \tau}{\partial z_i}\left(z_i ; \mathbf{h}_{i}\right)\right|
\]

However, the diffeomorphism constraint imposed on the transformer is highly restrictive, requiring it to be a monotonic and $\mathcal{C}^1$ function. To enhance expressiveness, various flows have been developed. In the upcoming section, we will use Neural Spline Flows, specifically focusing on Rational-Quadratic Neural Spline Flows, which are among the most expressive flows developed to date \cite{durkan2019neural}.

\subsection{Neural Spline Flows}
\label{Neural Spline Flows}

Significant efforts have been dedicated to a specific class of normalizing flows known as Neural Spline Flows (NSF) \cite{durkan2019neural}.  Splines are piecewise-defined differentiable functions. In the context of NSFs, each transformer is a spline, with each piece serving as a bijective and differentiable function on its defined segment. To guarantee that the transformer remains bijective and differentiable, making it a $\mathcal{C}^1$ function, the overall spline should not only be monotonic, but also maintain continuity along with its derivative. 

\subsubsection{Rational-Quadratic Neural Spline Flows}

Durkan et al. \cite{durkan2019neural} advocated for the use of monotonic rational-quadratic spline flows (RQ-NSF), where the transformers are rational-quadratic functions, i.e. the ratio of two quadratic functions. Rational-quadratic splines have a convenient flexibility due to their infinite Taylor-series expansion while being defined by a small number of parameters. Additionally, these splines are analytically invertible and differentiable. 

The overall spline acts as a diffeomorphism, providing a smooth one-to-one mapping within a specific region of interest, typically chosen as the segment $[A, B]$. Within this segment, the transformer distorts the parameter space, while beyond this interval, it is the identity. This transformation is achieved through the use of the rational-quadratic splines parametrization introduced by Gregory and Delbourgo \cite{delbourgo1983c}. The parametrization involves a total of N different rational-quadratic functions, with their boundaries determined by pairs of coordinates ${\left(x^{(n)}, y^{(n)}\right)}_{n=0}^N$ known as knots, where the spline passes through. To ensure continuity, the first knot is $\left(x^{(0)}, y^{(0)}\right)=(A,A)$ and the last knot is $\left(x^{(N)}, y^{(N)}\right)=(B, B)$. In order to parameterize a monotonic spline, N-1 intermediate positive derivative values ${(f^{(n)})}_{n=1}^{N-1}$ need to be defined. The derivatives at the boundary points are set to 1 to match the identity function (i.e., $f^{(0)}=f^{(N)}=1$). 

 \subsubsection{Masked Autoregressive Network}

The conditioners $\mathbf{h}_i$ associated with the transformers, responsible for providing the parameters of the RQ-NSF are expressed as functions of the autoregressive input features $\mathbf{z}_{<i}$. However, implementing these conditioners as separate neural networks for each dimension is computationally inefficient, particularly for high-dimensional data. 

To overcome this issue, a solution known as masked autoregressive network (MAN) \cite{germain2015made} is adopted in this work. The masked autoregressive network takes the entire vector $\mathbf{z}$ as input and directly outputs all the parameters of the $D$ conditioners $\left(h_1, h_2, \ldots, h_D\right)$ simultaneously. This is achieved by modifying a standard feed-forward network, to ensure that no connections exist from the input $z_{\geq i}$ to the outputs $\left(h_1, \ldots, h_i\right)$. This connection cut is implemented by element-wise multiplication of the weight matrices connecting the neurons of the network with a binary matrix of the same size. The binary matrix "masks out" the undesired connections by setting them to zero, preserving all other connections. This network was used by Papamakarios et al. \cite{papamakarios2018masked}  to parameterize flows, leading to masked autoregressive flows (MAF). 

However, this can lead to richer relations for later components and poorer relations for the first dimensions. To ensure that all input variables interact with each other, random permutations of dimensions are commonly introduced between autoregressive flows. 

\subsection{Normalizing Flows for Amortized Bayesian Inference}
\label{Normalizing Flows for Amortized Variational Inference}

The previous section highlighted the advantages of NFs for density estimation due to their expressive nature. Using the previous definition, NFs are effective at estimating a density $p_\theta(\mathbf{x})$. Building on this, a natural question arises: can we extend NFs to learn conditional probabilities $p_\theta(\mathbf{x}|\mathbf{z})$ in a similar manner, where $\mathbf{z}$ is a condition provided by the user at inference time? We will keep the notation of Section \ref{problem definition} with the input dataset $\mathbf{X}_d$ from where we want to infer the target posterior distribution, $p(\beta|\mathbf{X}_d)$.

A commonly used approach in this context is to apply Normalizing Flows to Variational Inference (VI), as noted in previous works \cite{rezende15,berg2019sylvester}. VI requires evaluating the likelihood. In our study, the computation of the Poisson likelihood is feasible, but extremely time and computationally expensive at training time and sampling time, especially for high dimensional $\mathbf{X}_d$ or $\beta$. Addressing this limitation, Likelihood-free Inference (LFI) emerges as an alternative. LFI refers to a set of methods used for statistical inference when the likelihood function (or equivalently the target posterior distribution) is either unavailable or too computationally intensive to calculate.  The potential of Normalizing Flows for LFI has been previously explored in works like \cite{papamakarios2019sequential,winkler2023learning}. 

In this work, we investigate a similar method as in \cite{papamakarios2019sequential} utilizing Normalizing Flows. This method bypasses the need for direct likelihood evaluation during training and sampling, and instead train the model using samples of the target posterior distribution that have already been generated. We generate samples of the target posterior distribution using a simulation process detailed in Appendix \ref{Generating the datasets}, which serve as the basis for computing our loss function, as described in Equation \ref{Dkl gradient}. These samples can be generated once prior to training. Another key feature of our model is its amortization aspect, which enables the inference process to generalize across different \( \mathbf{X}_d \) datasets, thereby eliminating the need for separate training for each dataset. This model approximates the posterior distribution of latent variables for a wide range of \( \mathbf{X}_d \) datasets. At the heart of our model are one or more encoder neural networks designed to distill information from the conditional variable \( \mathbf{X}_d \) into the parameters of the flow transformation.

\begin{figure}[h!]
\centering
    \includegraphics[width = 0.48\textwidth, trim=0cm 0cm 0cm 0cm, clip]{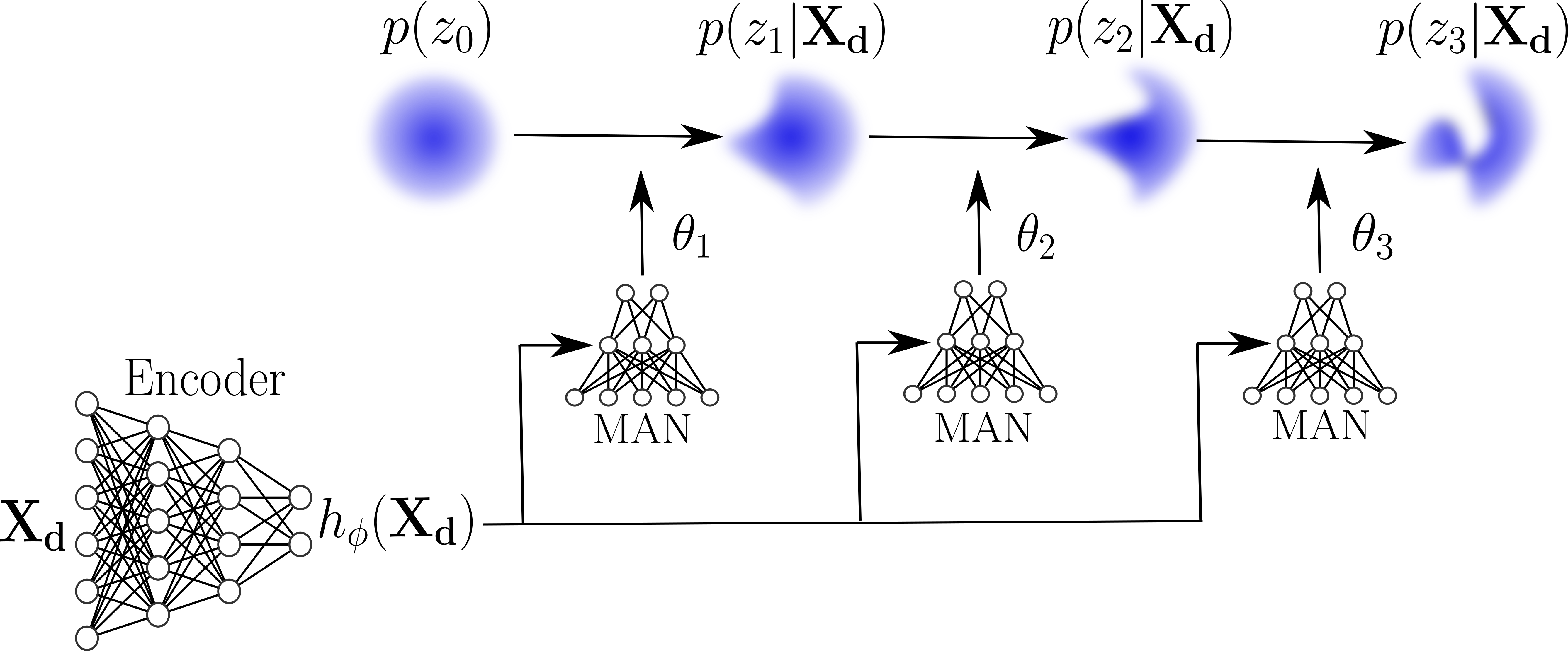}
    \caption{Concept of a conditional normalizing flows model using masked autoregressive networks (MAN). The encoder network outputs a lower dimension representation of the input dataset which is fed into the MAN of each NF, producing input-dependent flows parameters $\theta_i$.  }
    \label{fig: ANF}
\end{figure}

NF models embedding encoder networks are referred to as Conditional Normalizing Flows (CNF) in this work. Such models are particularly used in Amortized Variational Inference \cite{rezende15,berg2019sylvester}. CNFs can be seen as estimators of a family of flows, instead of a single flow, with the ability to choose the right flow under the conditions given by the input. The encoder networks learn a lower-dimensional representation $h_\phi(\mathbf{X}_d)$ of the dataset, referred to as context features, where $\phi$ are the parameters of the encoder network, and their outputs are fed into each MAN of the Normalizing Flows. This results in an input-dependent flow transformation, endowing it to be used for Amortized Bayesian Inference as it will be applied in the following section. An illustration of the concept of CNF is given in Figure \ref{fig: ANF}.
\section{Conditional Normalizing Flows for the Near Detector fit at T2K}
\label{Amortized Normalizing Flows for the Near Detector fit at T2K}

In this section, CNF will be used to sample from the posterior probability $p(\beta | \mathbf{X}_d)$ given a dataset $\mathbf{X}_d$. The implementation of RQ-NSF is based on the \texttt{nflows} Pytorch implementation of Durkan et al.  \cite{nflowsgit}. In this work, the training is done in two steps, which will be detailed in Section \ref{Training methodology}. Section \ref{Model's architecture and training} delves into the architecture of the CNF model. Section \ref{Model's performances} presents the performance of the model tasked with the prediction of the posterior distribution accounting for Poisson fluctuations at inference time. In our exploratory methodology, we have simplified the problem by segmenting the neutrino flux into three energy bins, each associated with a reweight variable \( \beta \) ranging from \( 0.5 \) to \( 1.5 \). Consequently, the latent space under consideration is the cube \( [0.5, 1.5]^3 \).

\subsection{Training methodology }
\label{Training methodology}

Training the model involves simultaneously learning two essential components: (1) the context features, here generated by two encoder networks, and (2),  RQ-NSF transformations. 

\begin{figure}[htbp!]
\centering

\begin{minipage}{0.165\textwidth}
  \centering
  \includegraphics[width=0.8\linewidth]{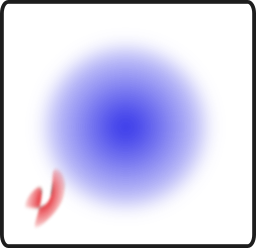}
  \subcaption{}

\end{minipage}%
\begin{minipage}{0.165\textwidth}
  \centering
  \includegraphics[width=0.8\linewidth]{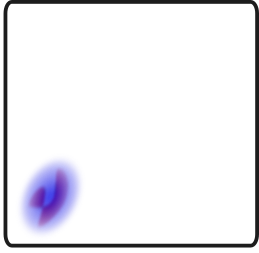}
  \subcaption{}
\end{minipage}%
\begin{minipage}{0.165\textwidth}
  \centering
  \includegraphics[width=0.8\linewidth]{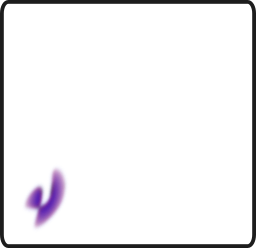}
  \subcaption{}
\end{minipage}

\caption{Representation of the predicted posterior distribution during training. The current approximation of the posterior distribution is represented in blue and the target posterior distribution  in red. (a) corresponds to the initial state, (b) to the transition between Step $\mathbf{1.}$ and Step $\mathbf{2.}$, and (c) to the end of the training.}
\label{fig:ANFconcepttrain}
\end{figure}

The training is divided into two steps. Figure \ref{fig:ANFconcepttrain} shows a conceptual representation of the posterior distribution learning during these two phases.

$\mathbf{1.}$ In the initial stage, the objective is to obtain an approximate representation of the posterior distribution using only a simple linear flow, the RQ-NSFs being frozen and initialized as the identity. The aim is to convert a normally distributed variable $\mathbf{u} \sim \mathcal{N}(0,\mathbf{I})$ into a tridimensional correlated Gaussian variable $\mathbf{x} \sim \mathcal{N}(\mu,\Sigma)$ that closely resembles $p(\beta | \mathbf{X}_d)$. Given a normal variable $\mathbf{u}$, we can simply add a shift and correlations with a linear transformation:
$$\mathbf{x}= L\mathbf{u} +\mu.$$
Here, $L$ corresponds to the Cholesky decomposition of the covariance matrix $\Sigma = L L^T$, and $\mu$ represents the mean of the $\mathbf{x}$ distribution. Both the Cholesky matrix and the mean are learned and directly output by the first encoder network only, and we note its parameters as  $\psi$. The only requirement for this linear flow is that $L$ is a lower-triangular matrix with strictly positive diagonal elements. This first step is important for two reasons.  Firstly, it prevents training instabilities by initially shifting and scaling $p(\mathbf{u})$ to cover $p(\beta | \mathbf{X}_d)$. More crucially, it empowers the complex flows to fully use their expressiveness in learning local non-Gaussian characteristics, rather than expending their potential on locating and scaling $\mathbf{u}$.

During this step, the loss function can be expressed as :
$$
\begin{aligned}
L(\psi) & = D_{\mathrm{KL}}\left(p(\beta | \mathbf{X}_d) \| q_{\psi}(\beta | \mathbf{X}_d)\right) \\
& \approx \frac{1}{N} \sum_{j=1}^N \sum_{i=1}^3 [\frac{1}{2}(L^{-1}(\beta_j-\mu))_i^2 + \log(\sigma^{i})]
\end{aligned} 
$$
Here, $\sigma^i$ corresponds to the i-th diagonal element of $L$.   To maintain clarity, we have purposefully refrained from introducing an additional summation that would account for the mean loss expectation across various datasets $\mathbf{X_{d}}$.

The first term within this expression bears a strong resemblance to a $\chi^2$ term, and it reaches minimal values when $\mu$ is equal to the mean of the $\beta$ samples and when the introduced spreads $\sigma^i$ from $L$ are maximized. The second term is proportional to the Shannon entropy of the predicted posterior distribution, exerting an opposing effect to the first term. In particular, the reduction in entropy leads to a minimization of $\sigma^i$. 

$\mathbf{2.}$  After 10$\%$ of the training, we enable the RQ-NSFs, before the linear flow.  We learn context features noted $h_\phi(\mathbf{X_d})$ from a second encoder network with parameters $\phi$. $h_\phi(\mathbf{X_d})$ is fed into the MANs of the RQ-NSFs which have their own parametrization $\theta$. These flows introduce input-dependent non-Gaussian characteristics to the predicted posterior distribution $q_{(\psi,\theta,\phi)}(\beta | \mathbf{X}_d)$. To ensure smooth transitions in the loss function between the stages $\mathbf{1.}$ and $\mathbf{2.}$, we initialize the RQ-NSFs to the identity transformation. Consequently, the expression for the predicted $\beta$ can be written as:
$$\mathbf{\beta}= L[\mathbf{T_K \circ} ... \circ T_1 (\mathbf{u}) ]+\mu$$
Upon enabling the additional flows, the loss function transforms into:
\begin{align}
L(\psi,\theta,\phi)=& \frac{1}{2N} \sum_{j=1}^N \sum_{i=1}^3  [T_{K}^{-1} \circ ... \circ T_{1}^{-1}(L^{-1}(\beta_j-\mu))]_i^2 \notag\\
& + \sum_{i=1}^3[\sum_{k=1}^K\log (|J_{T_{k}}|_{ii}) +\log(\sigma^{i})] \notag
\end{align}

The complexity of this loss function may seem greater compared to the previous one, yet it can still be broken down into a combination of a $\chi^2$ term and an entropy term. What is important to note is that this formulation also reveals the computational efficiency of the backward computations, which exhibit linear complexity concerning both the number of flows and the dimensions of the posterior distribution. This property enhances the training's scalability, allowing it to be effectively applied to higher dimensions with larger numbers of flows.

\subsection[Model's architecture and training]{Model's architecture}
\label{Model's architecture and training}

\begin{figure}[htbp!]
\centering
    \includegraphics[width = 0.45\textwidth]{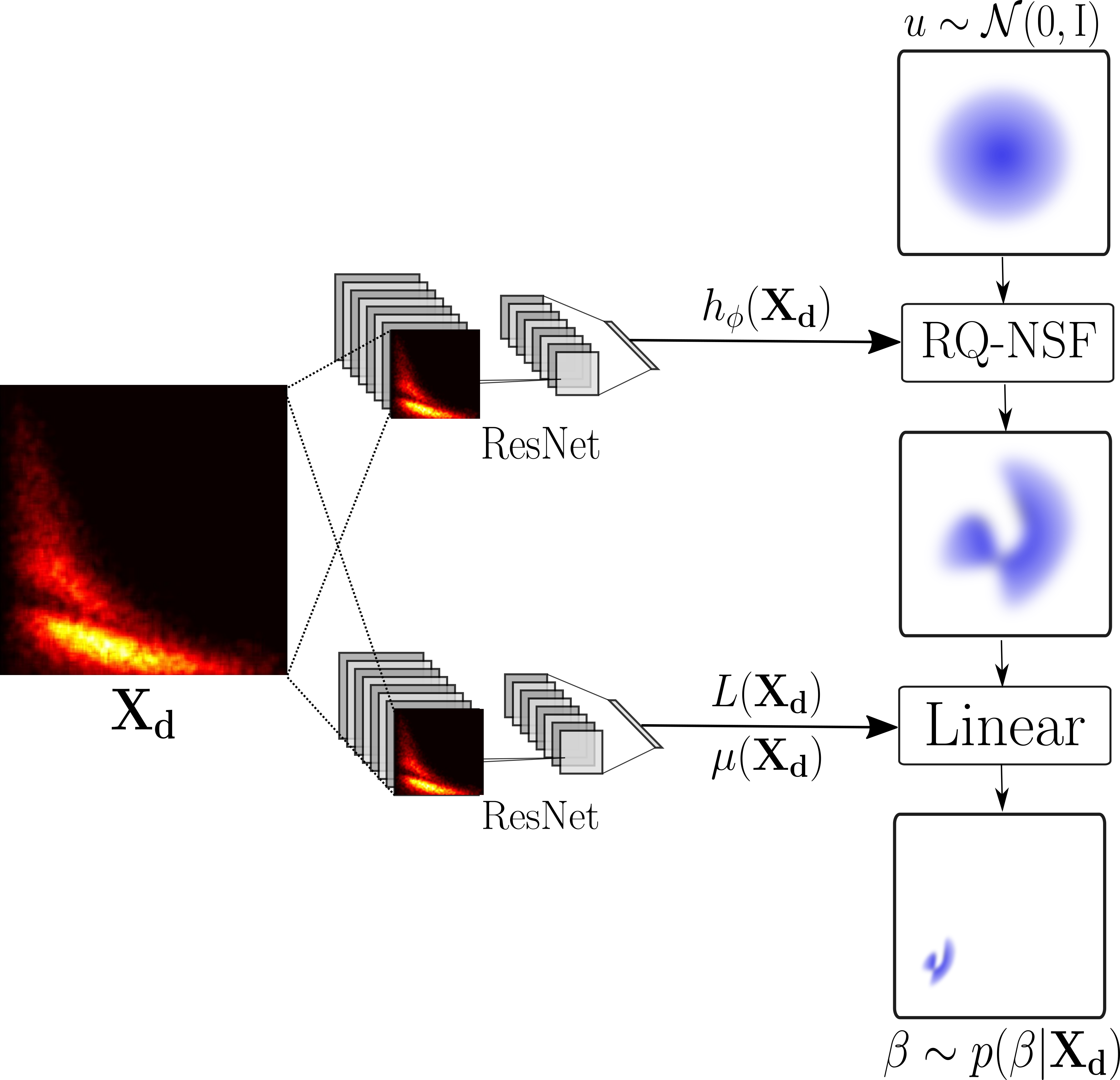}
    \caption{Model's architecture used for the inference of $p(\beta | \mathbf{X}_d)$ from a dataset $ \mathbf{X}_d$}
    \label{fig: ANF scheme}
\end{figure}

The architecture of the model is represented in Figure \ref{fig: ANF scheme}. The model's structure can be broken down into two main blocks: the Encoder block and the Flows block encompassing the RQ-NSFs and the Linear flow. The Encoder block contains two encoder networks, both utilizing a Convolutional Neural Network (CNN) architecture, chosen for its capacity in extracting information from image-based data. This feature is particularly beneficial for processing our bidimensional histogram inputs, $\mathbf{X_d}$. One CNN yields both the Cholesky Matrix $L$ and $\mu$, while the second CNN generates the context features used by the RQ-NSFs. Both CNNs share a common architecture, differing only in their last fully connected layers. The architecture is based on a fine-tuned ResNet-50 model \cite{he2015deep} followed by one fully connected layer.  The first CNN produces 9 output nodes (6 for $L$ and 3 for $\mu$), while the second CNN generates 10 context features. Although the number of context features proved sufficient for the study, it is important to note that a complex distribution may require more context features.

\noindent The initial phase of the flow model involves a sequence of K=4 blocks, each comprising an RQ-NSF followed by a generalized LU permutation as defined by Oliva et al. \cite{pmlr-v80-oliva18a}. The parameters of the RQ-NSFs are learned through a MAN consisting of three hidden layers, each containing 256 nodes. The MAN outputs the parameters of a 9-knot tri-dimensional spline. The linear flow is only parameterized by the output of the first CNN.  

The model is trained using $7,500$ $\mathbf{X_d}$ datasets of muon observables, and for each dataset, $4,000$ samples from the target posterior distribution to compute the loss as in  Equation \ref{Dkl gradient}. The generation of the datasets and the posterior samples is detailed in Appendix \ref{Generating the datasets}.

\subsection{Model's performances}
\label{Model's performances}

\begin{figure*}[ht!]
\centering
    \subcaptionbox{Target posterior distribution}
    [.49\linewidth]{\includegraphics[width = 0.99\linewidth, trim=0cm 0cm 0cm 0cm, clip]{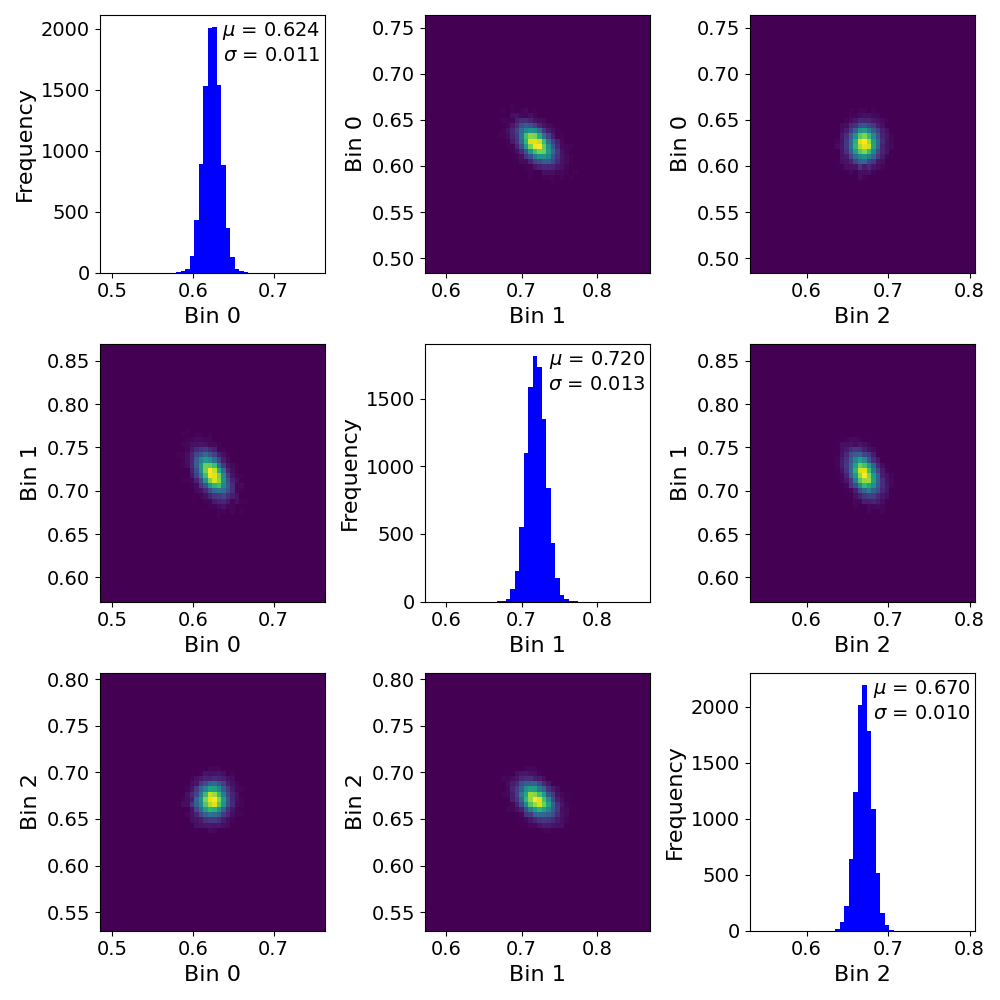}}
    \subcaptionbox{Predicted posterior distribution}
    [.49\linewidth]{\includegraphics[width = 0.99\linewidth, trim=0cm 0cm 0cm 0cm, clip]{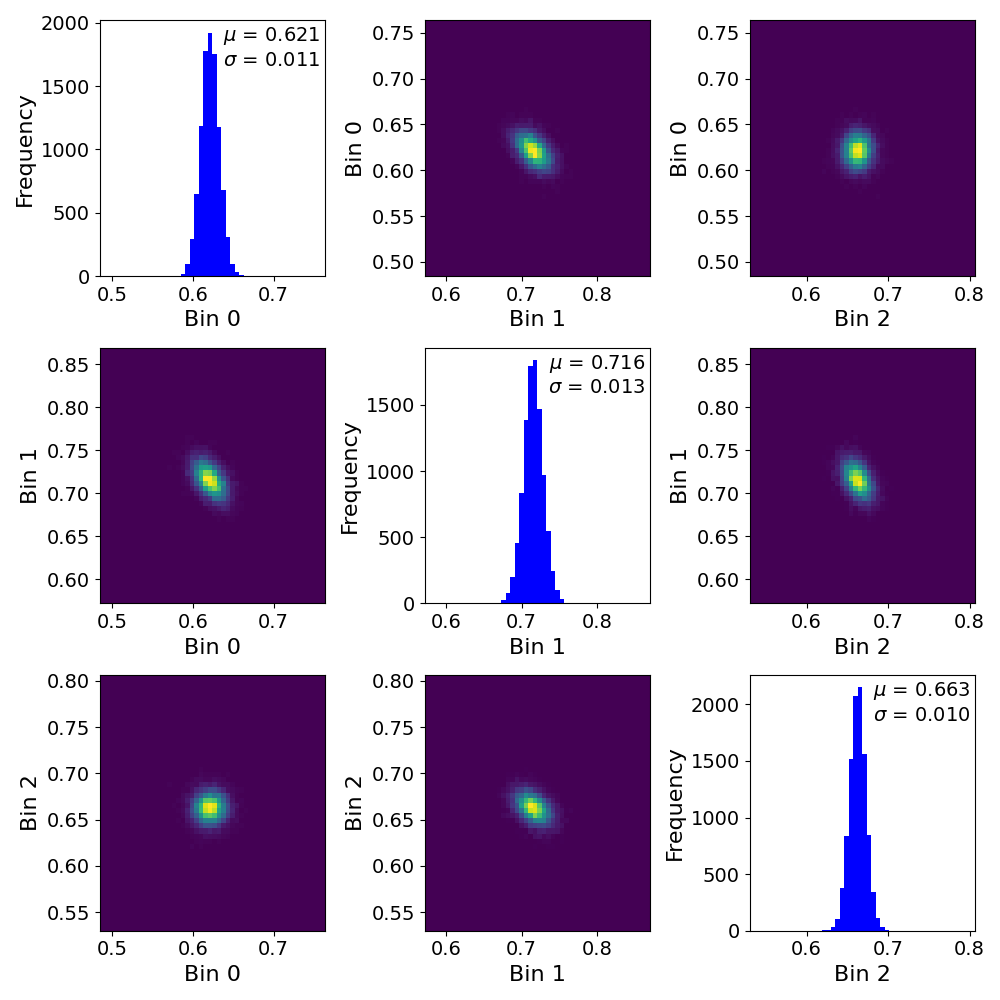}}
    \caption{10,000 Samples from the target posterior distribution (a) and from the predicted posterior distribution (b). The diagonal plots represent the marginal distribution of the 3 reweights bins $\beta_i$. The off-diagonal plots correspond to the 2D histograms of $(\beta_i,\beta_j)$.}
    \label{fig: ANF Poisson prediction}
\end{figure*}

The present section focuses on evaluating the model’s performances up to the fourth moment prediction for the posterior distribution. In this analysis, we compare two posterior distributions: the sampled predicted posterior distribution using NF \( q_\text{NF}(\beta | \mathbf{X_d}) \) and the sampled target distribution \( p_\text{target}(\beta | \mathbf{X_d}) \) used to compute the loss during training along with the designed reweight \(\beta_\text{Asimov}\), referred to as the Asimov datapoint, used to generate \( \mathbf{X_d}\).  \( p_\text{target}\) is produced by generating Poisson fluctuations of $\mathbf{X_d}$ and identifying the most likely reweight for each fluctuation. Therefore, both \( p_\text{target}\) and \( q_\text{NF}\) retains the initial bias inherent in the creation of the dataset $\mathbf{X_d}$. Consequently, we focus on comparing \( q_\text{NF}\) not with \( \beta_\text{Asimov}\) but with \( p_\text{target}\). 

\begin{figure}[h!]
\centering
    \includegraphics[width = 0.42\textwidth, trim=0cm 0cm 2cm 0cm, clip]{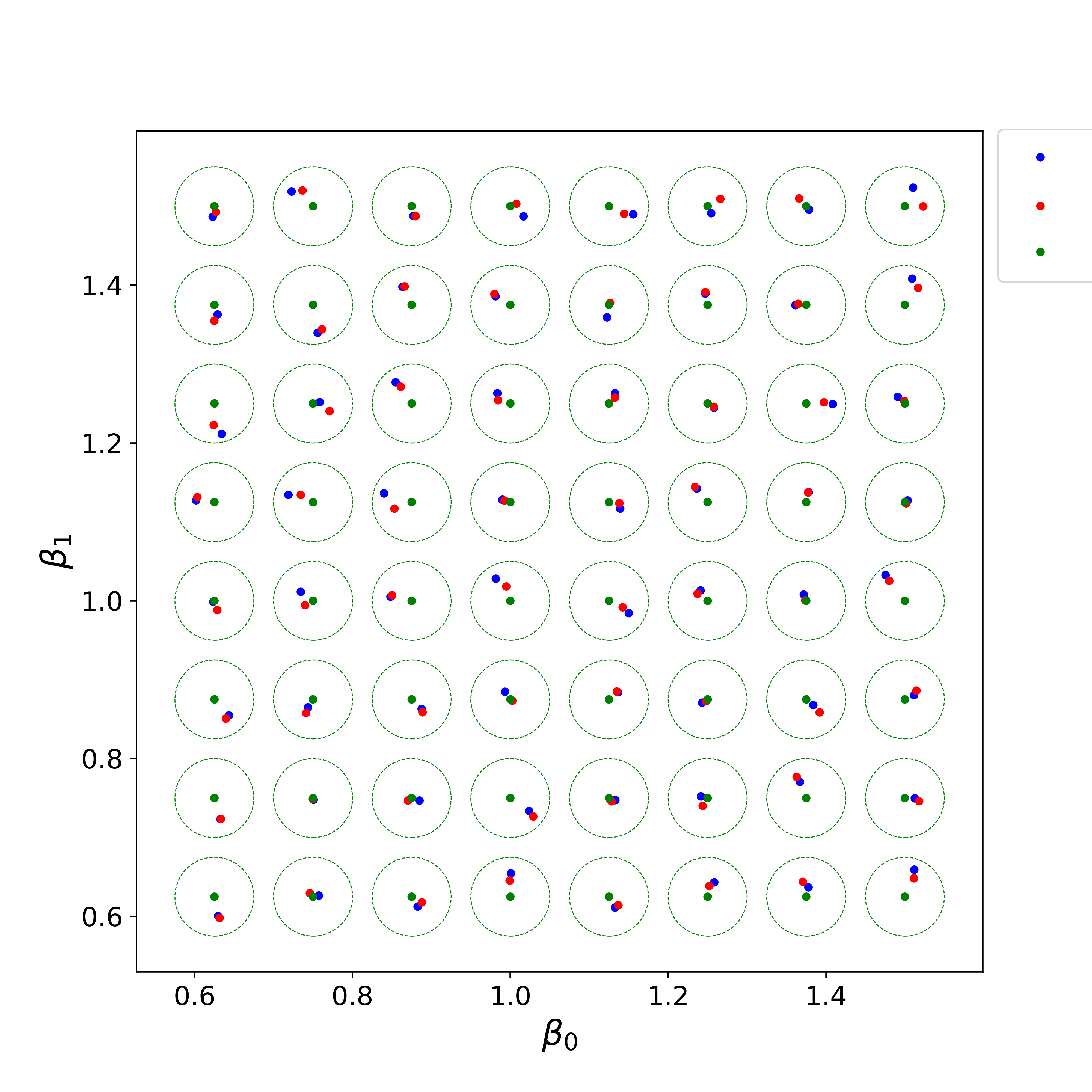}
    \caption{Projection in the $(\beta_0,\beta_1)$ plane of the mean of the predicted posterior distribution (blue), the reweight given $\mathbf{X}_d$ using likelihood maximization (red) and the Asimov datapoints (green) on a grid and for a constant $\beta_2$ of 1. The circles in dashed lines have a radius of 0.05 and are centered at the designed $\beta$ value.}
    \label{fig: ANF grid}
\end{figure}

For each \(\beta_\text{Asimov}\), we generate a dataset $\mathbf{X_d}$ from which we generate 10,000 samples from \( q_\text{NF}(\beta | \mathbf{X_d}) \) and 10,000 samples from \( p_\text{target}(\beta | \mathbf{X_d}) \). A comparison between \( q_\text{NF}(\beta | \mathbf{X_d}) \) and \( p_\text{target}(\beta | \mathbf{X_d}) \) is shown in Figure \ref{fig: ANF Poisson prediction} for a specific dataset $\mathbf{X_d}$. The model seems able to predict correctly the shape of the posterior distribution including correlations and higher order moments. To evaluate the model’s performance, we analyze its predictions over a complete range of \( \beta \) values. Figure \ref{fig: ANF grid} presents this analysis, offering a comparison among the means of \( q_\text{NF}(\beta | \mathbf{X_d}) \), \( p_\text{target}(\beta | \mathbf{X_d}) \), and the Asimov datapoint.

Additionally, we quantify the  accuracy using the coefficient of determination, denoted as \( R^2 \) for $N=1,000$ Asimov datapoints uniformly sampled from the $[0.5,1.5]^3$ cube. In this case, \( R^2 \) is defined as:

\[ R^2 = 1 - \frac{\sum_i||\beta_\text{Asimov} - \hat{\beta}||^2}{\sum_i||\beta_\text{Asimov} - 1||^2}=1-\frac{12}{3N} \sum_i||\beta_\text{Asimov} - \hat{\beta}||^2 \]

where \( \hat{\beta} \) represents the mean of the \( q_\text{NF}\) (or \( p_\text{target}\)) and \( \beta_\text{Asimov}\) denotes the Asimov datapoint. An \( R^2 \) of 1 implies that the model predicts the mean with perfect accuracy, while an \( R^2 \) of 0 indicates that the model randomly predicts the mean. For our model, the \( R^2_\text{NF} \) value is 0.9975. This score is marginally lower than the  \( R^2_\text{target} \) of 0.9980 using \( p_\text{target}\). The  coefficient of determination is further reflected in the Root Mean Square (RMS) errors: $0.013$ when comparing the mean of \( p_\text{target}\) against the Asimov data point, and $0.014$ for the mean of \( p_\text{NF}\) against the Asimov data point.  This highlights that the model's deviations are minor compared to the Poisson fluctuations in generating the dataset \(\mathbf{X}_d\).

Our goal is to assess the alignment of \( q_\text{NF}\) with \( p_\text{target}\), particularly in terms of the first four statistical moments: mean, standard deviation, Pearson correlation factors, skewness, and kurtosis. The discrepancies in these statistical moments come from two sources: (1) statistical variations due to the finite sample size from posterior distributions, and (2) systematical discrepancies introduced by the model itself. A comprehensive analysis of these discrepancies for the first statistical moments is detailed in Table \ref{tab:my-table} within Appendix \ref{Statistical and systematical errors in the statistical moment predictions}. Notably, it is demonstrated that both distributions closely approximate multivariate Gaussian distributions.

In our study, we introduce a metric designed to quantify the shape differences between the two distributions, \( q_{\text{NF}} \) and \( p_{\text{target}} \), across various Asimov data points. To achieve a balanced measure of these discrepancies, we employ a symmetrical version of the Kullback-Leibler divergence:

$$\mathbf{D}_\text{s}(q_\text{NF}, p_\text{target}) = \frac{D_\text{KL}(q_\text{NF}|| p_\text{target}) + D_\text{KL}( p_\text{target}||q_\text{NF})}{2}.$$

We estimate \(\mathbf{D}_{\text{s}}\) using a Monte Carlo estimate, as outlined in Equation \ref{KL}, for datasets of \( q_{\text{NF}} \) and \( p_{\text{target}} \) with equal sample sizes, and we average this over 200 Asimov data points. To focus solely on shape discrepancies, we remove the mean from both datasets.

Furthermore, we compare the expected value of $\mathbf{D_s}$ between \( q_{\text{NF}} \) and \( p_{\text{target}} \) by examining the metric in a simpler context: two datasets derived from centered three-dimensional correlated Gaussian distributions with different covariance matrices. We generate two datasets of $N$ samples respectively from $\mathcal{N}(0, \Sigma)$ and $\mathcal{N}(0, \Sigma + \Delta\Sigma)$. Here, $\Sigma$ is the average covariance matrix in our study (as detailed in Table \ref{tab:my-table}). The covariance difference, $\Delta\Sigma$, is a random symmetric matrix. We derive its elements by drawing $\Delta L_{ij}$ from $\text{Unif}([-\sqrt{3}\alpha|L_{ij}|, \sqrt{3}\alpha|L_{ij}|])$ for $j\leq i$, where $L$ is the Cholesky decomposition of $\Sigma$, and $\Delta L$ represents the deviation in the Cholesky matrix between the two distributions. The parameter $\alpha$ represents the root mean square relative mismatch (RMSRM) in the Cholesky matrix components between the two multivariate Gaussian distributions.

\begin{figure}[h!]
\centering
    \includegraphics[width = 0.48\textwidth, trim=1cm 0cm 1cm 2cm, clip]{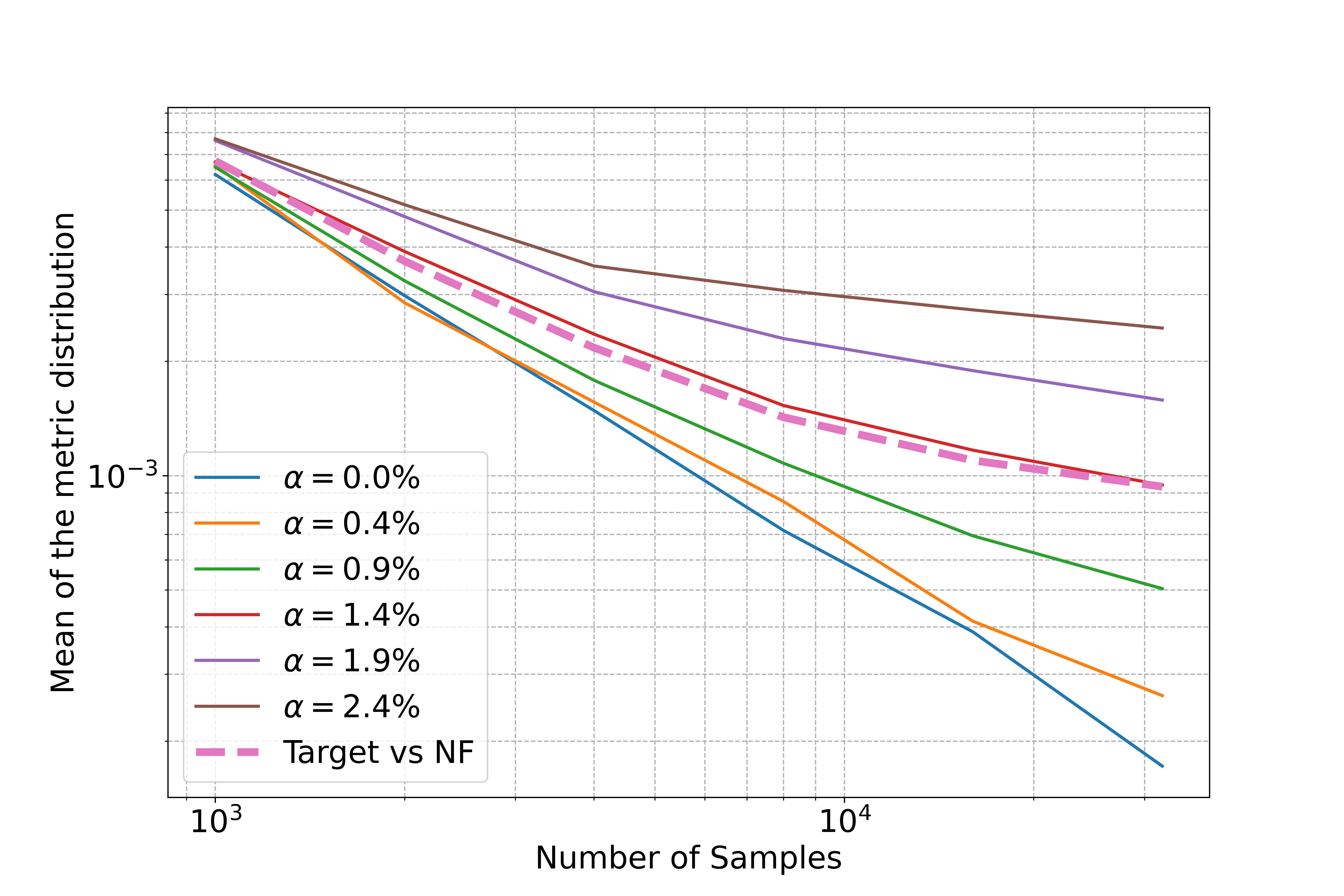}
    \caption{Evolution of $\mathbf{D_s}$ with sample size for different values of $\alpha$. The evolution of $\mathbf{D_s}$ using \( q_\text{NF}\) and \( p_\text{target}\) is represented with the thicker dashed line.}
    \label{fig: Cov mismatch}
\end{figure}

Figure \ref{fig: Cov mismatch} illustrates the evolution of $\mathbf{D_s}$ with the sample size for different values of $\alpha$. We also show the evolution of $\mathbf{D_s}$ for \( q_\text{NF}\) and \( p_\text{target}\). It shows the importance of statistical errors in the predictions since the metric \(\mathbf{D}_{\text{s}}\) shrinks drastically with the sample size. Notably, the trend of $\mathbf{D_s}$ with increasing sample size using \( q_\text{NF}\) and \( p_\text{target}\) corresponds with that of the mismatched Gaussian distributions at $\alpha=0.014$. This implies that the RMSRM in the standard deviations between the target and predicted distributions is approximately $1.4\%$, which is in agreement with Table \ref{tab:my-table}.

Furthermore, this inference method offers advantages in terms of sampling speed while keeping a high predictive accuracy. Traditional methods based on likelihood estimation are inherently limited in speed by the process of estimating the likelihood. For instance, calculating the likelihood as defined in Equation \ref{likelihood equation} a million times on one of our CPU cores requires $1135$ seconds. Consequently, for approaches that need at least one likelihood estimation for each posterior sample (like the Metropolis-Hasting algorithm used in MCMC), the fastest possible sampling rate assuming a 100$\%$ sampling efficiency is a million posterior samples in $1135$ seconds. In contrast, the model’s sampling speed represents a more efficient alternative, generating 1 million samples from \( q_\text{NF}\) in $174$ seconds on the same CPU core. This translates to a minimum improvement in sampling speed by a factor of $6.5$. Such an increase in speed during inference is particularly advantageous for real-time applications, including object tracking or stock price prediction.   This is particularly interesting if the model can predict non-gaussian posterior distribution. This question will be addressed in the following section.

\begin{figure*}[htbp!]
\centering
    \subcaptionbox{Target posterior distribution}
    [.49\linewidth]{\includegraphics[width = 0.99\linewidth, trim=0cm 0cm 0cm 0cm, clip]{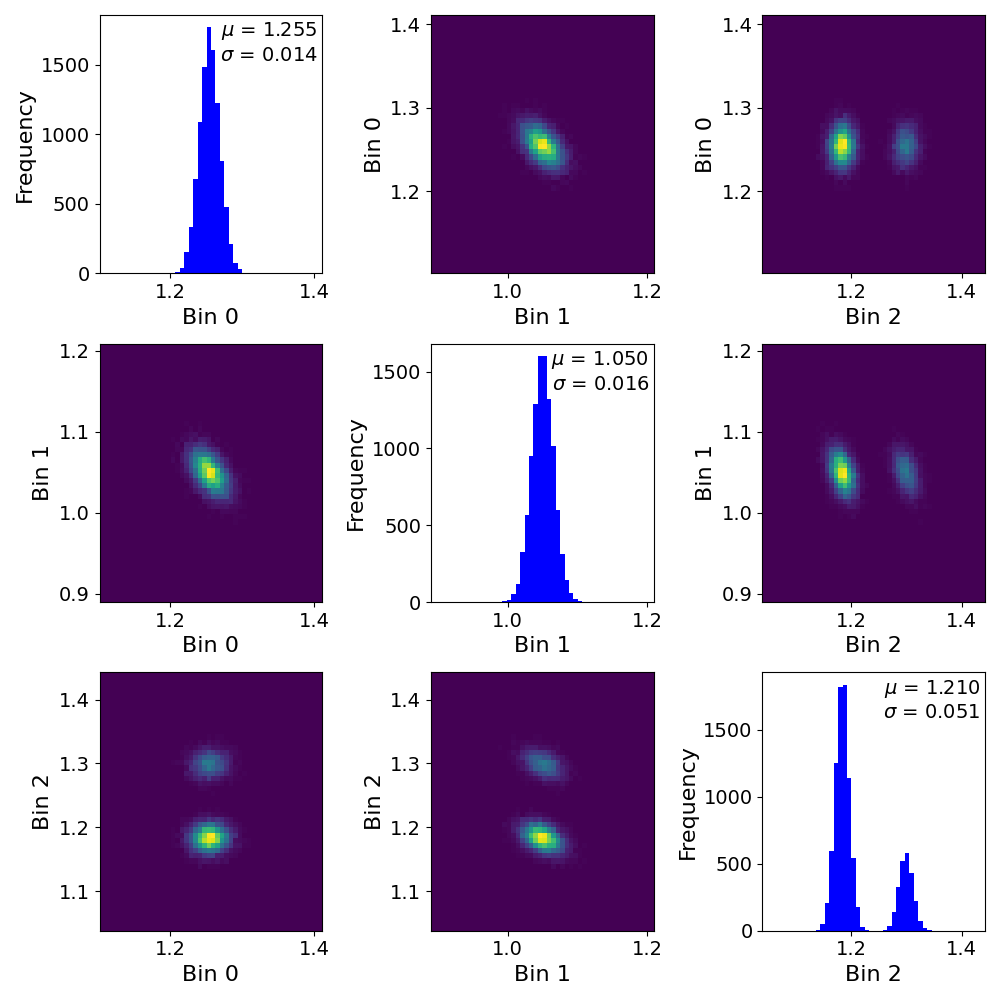}}
    \subcaptionbox{Predicted posterior distribution}
    [.49\linewidth]{\includegraphics[width = 0.99\linewidth, trim=0cm 0cm 0cm 0cm, clip]{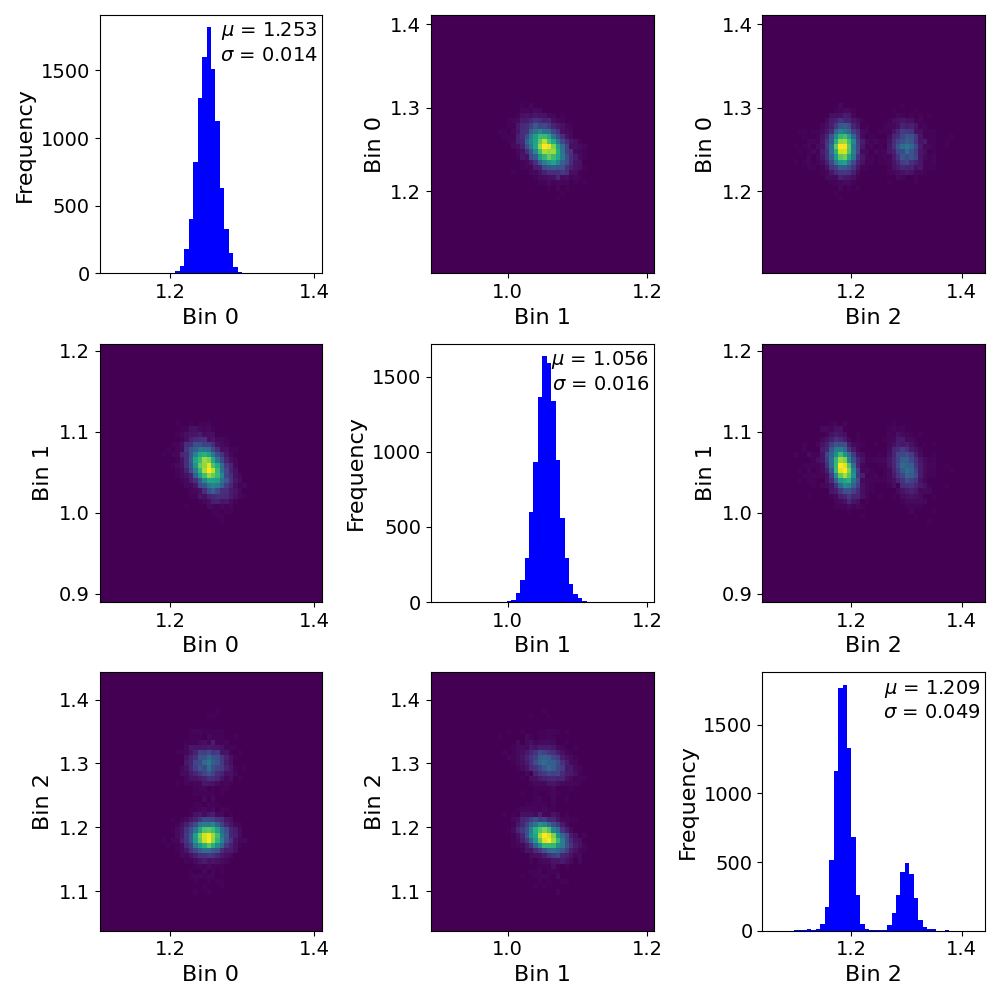}}
    \caption{10,000 Samples from the target posterior distribution (a) and from the predicted posterior distribution (b). The diagonal plots represent the marginal distribution of the 3 reweights bins $\beta_i$. The off-diagonal plots correspond to the 2D histograms of $(\beta_i,\beta_j)$.}
    \label{fig: ANF Poisson prediction island}
\end{figure*}

\begin{figure}[h!]
\centering
    \includegraphics[width = 0.42\textwidth, trim=0cm 0cm 0cm 0.7cm, clip]{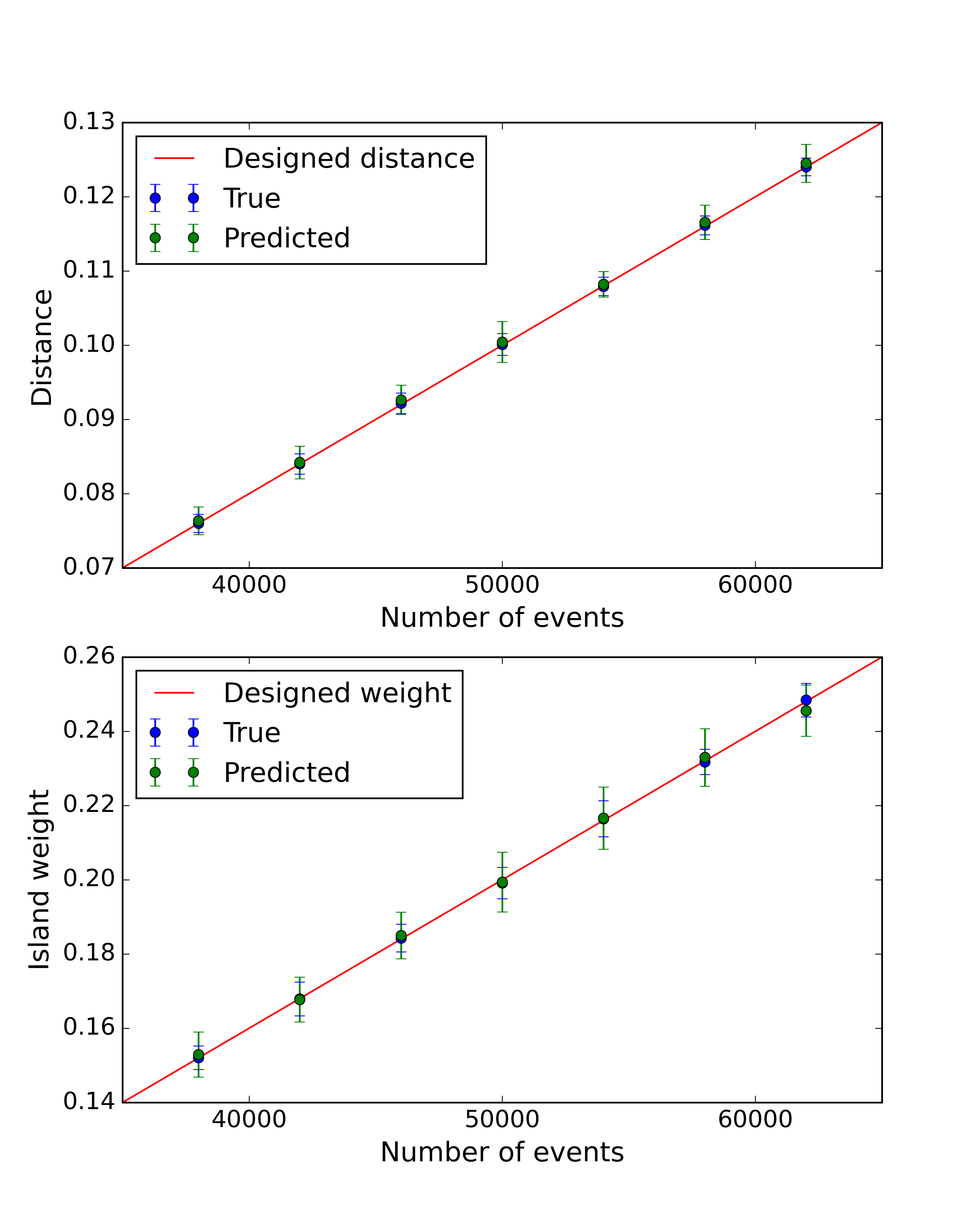}

    \caption{Variation of the distance between modes (top) and the island weight (bottom) with the event counts. Error bars represent $\pm 1 \sigma $ deviations calculated for $50$ posterior distributions from $50$ different $\mathbf{X_d}$.}
    \label{fig: height shift}
\end{figure}

\section{Quantifying and Modeling Non-Gaussian Characteristics}
\label{Quantifying and Modeling Non-Gaussian Characteristics}

While the model demonstrated its proficiency in accurately predicting a posterior distribution, the last section did not extensively highlight the expressiveness of the RQ-NSF. The posterior distributions, accounting for the Poisson statistics, are close to multivariate Gaussian distributions. In particular, the important features, the mean, and the covariance matrix could potentially be learned by the linear flow only.

This section is an exploration of the model's behavior when tasked with predicting non-Gaussian attributes. To this end, we introduce modifications to \( p_\text{target}\) introducing a distinct "island" within the third energy bin dimension. This bimodal structure of the posterior distribution serves as a meaningful test case, as conventional methods like Metropolis Hasting's algorithm for Markov Chain Monte-Carlo often struggle to infer multimodal distributions. 

A crucial aspect of our model is put to the test: its capability to predict non-Gaussian features that depend on the input dataset $\mathbf{X}_d$. This test demonstrates that the context features learned from $\mathbf{X}_d$ and fed to the MAN influence the geometry of the \( q_\text{NF}\). To this end, the introduced shift and weight of the island are input-dependent and more precisely, proportional to the total event count in the dataset $\mathbf{X}_d$. 

This is done by taking the already generated samples from \( p_\text{target}\) and applying a stochastic transformation with probability $\textbf{p}$ of shifting by a quantity $\textbf{s}$ the posterior samples. We have chosen expressions for the probability and for the shift, defined as follows, where $\text{N}_\text{events}$ is the total event count in million:

\begin{equation*}
    \begin{aligned}
\textbf{p}(\text{N}_\text{events})=4\text{N}_\text{events}\\
\textbf{s}(\text{N}_\text{events})=2\text{N}_\text{events}
    \end{aligned}
\label{eq : shift height}
\end{equation*}

An example of the comparison of \( p_\text{target}\) and \( q_\text{NF}\) for a specific choice of reweight $\beta$ is given in Figure \ref{fig: ANF Poisson prediction island}. The close agreement between them illustrates the model's capacity to infer this non-gaussian characteristic. 

We aim to assess the model's accuracy in predicting the island weight (related to $\text{p}$) and the distance between the means of the two modes (related to $\text{s}$) across various $\text{N}_\text{events}$ values. We construct datasets for Asimov datapoint $\beta_\text{Asimov}$ corresponding to increasing event counts. We estimate the marginal distribution of $\beta_3$ by fitting the samples using a mixture of two Gaussian distributions for  \( q_\text{NF}\). We compute the distance between the modes using the absolute difference of the two means given by the fit while the weight of the island is directly given by the fit. 

Figure \ref{fig: height shift} presents the evolution of the predicted distance and island weight for $7$ different $\text{N}_\text{events}$ values between $37,500$ and $62,500$.  For each value of $ N_\text{events}$, the posterior distributions are predicted for $50$ $\beta_\text{Asimov}$ sampled in the hyperplane yielding $\mathbf{X_d}$ with  $ N_\text{events}$ events (see Appendix \ref{Generating the datasets} for more detail).  Although predictions for the island weight exhibit some variability, both the island weight and distance are consistently predicted with accuracy within a 1$\sigma$ margin from the designed values.

\section*{Conclusion}
In this work, we explored the potential of Normalizing Flows for Amortized Bayesian Inference in the context of the near detector fit at T2K. A brief introduction to Normalizing Flows as conditional density estimators is performed, emphasizing the particular implementation of Rational-Quadratic Neural Spline Flows (RQ-NSF). In a simplistic case, it was shown that such a model can predict accurately the posterior distribution of latent variables (the energy bin reweights) from observables (the muon momentum and angle) provided at inference time with a minimum improvement in sampling speed by a factor of $6.5$. In the last section, it was also demonstrated that Normalizing Flows can predict more intricate posterior distributions. However, the full potential of RQ-NSF might not have been fully tapped in this study. The exploration could extend to introducing multiple non-Gaussian attributes. In the context of the T2K experiment and neutrino-nuclei interactions, this method could also be extended with more work to more complex set of systematics. Ultimately, the demonstration of NF's flexibility and efficiency paves the way for more informed and accelerated Bayesian inferences, a prospect that holds substantial promise for a myriad of data-driven applications.

\section*{Acknowledgments}
The author acknowledges support from the Swiss National Foundation grants No. $200021E\_213196$.

\appendix
\section{Generating the datasets}
\label{Generating the datasets}

To generate the datasets, we divide the neutrino flux into 3 energy bins: [200 MeV, 620 MeV], [620 MeV, 800 MeV], and [800 MeV, 1.7 GeV]. For this study, we choose three equally populated bins using the nominal T2K flux where all systematics are set to their nominal values 1. The momentum \(p_\mu\) ranges from 0 to 1.7 GeV, the angle \(\theta_\mu\) ranges from 0 to \(\pi\). We choose a nominal distribution for the triplet $(p_\mu,\theta_\mu,E_\nu)$ corresponding to CCQE events generated by the NEUT Monte-Carlo generator \cite{Hayato:2009zz} with a "nominal" T2K neutrino flux at the near detector \cite{PhysRevD.87.012001}. 

\(\mathbf{\beta}\) is a tridimensional vector, where each component represents a reweight for a specific energy bin. Consequently, the probability of having a neutrino in the i-th energy bin is given by $ p(E_\nu^i|\mathbf{\beta})=\frac{\beta_i p(E_\nu^i)}{\sum\limits_{j}\beta_j p(E_\nu^j)}=\frac{\beta_i}{\sum\limits_{j}\beta_j}$.

Now, the modified joint probability of \(p_\mu\) and \(\theta_\mu\) becomes the sum of the individual probabilities corresponding to each energy bin:
\begin{eqnarray}
p(p_\mu, \theta_\mu|\mathbf{\beta}) & = \sum\limits_{i=1}^{3} p(p_\mu, \theta_\mu| E_\nu^i) \times p(E_\nu^i|\mathbf{\beta})  \notag \\ & =  \sum\limits_{i=1}^{3} p(p_\mu, \theta_\mu| E_\nu^i) \times \frac{\beta_i}{\sum\limits_{j}\beta_j}
\label{probability formula}
\end{eqnarray}
where \(E_\nu^i\) represents the neutrino energy corresponding to the i-th bin.

To train and test the model, we need multiple datasets, which are generated in the following way:

$\mathbf{1.}$ A sample of the reweight \(\mathbf{\beta}\), called Asimov datapoint, is taken from a uniform distribution in the range $[0.5,1.5]$ for each component.

$\mathbf{2.}$ The modified probability grid is computed using the Equation \ref{probability formula} multiplied by $\sum_{j}\beta_j$ to account for the increasing total number of events for increasing reweight values.

$\mathbf{3.}$ \((p_\mu, \theta_\mu)\) samples are sampled from this grid using the Accept-Reject Monte-Carlo technique. The total number of generated samples is proportional to $\sum_{j}\beta_j$ and goes from $25,000$ to $75,000$. This means that we train our model on datasets of varying size, where the datasets corresponding to the nominal reweight value $\beta=[1,1,1]$ have $50,000$ events. 

$\mathbf{4.}$ The \((p_\mu, \theta_\mu)\) events are stored in a histogram with a size of \(200 \times 200\), noted $\mathbf{X}_d$.

$\mathbf{5.}$ This process is repeated $7,500$ times for distinct Asimov datapoints \(\beta\), generating $7,500$ datasets containing between $25,000$ and $75,000$ \((p_\mu, \theta_\mu)\) samples. 

\begin{table*}
\caption{\label{tab:my-table}Table of the comparison of a chosen set of statistics between the predicted and target posterior distribution. The values are averaged across the 200 Asimov datapoints. $a$ and $b$ corresponds to the parameters of the fit of the MSE. $\overline{S_\text{target}}$ corresponds to the mean of the statistic from the sampled target posterior distribution. }
\begin{ruledtabular}
\begin{tabular}{cccccc}
\textbf{Bin} & \( \overline{S_\text{target}} \) & \textbf{a} & \textbf{b} & \textbf{Bias} ($\sqrt{\textbf{b}}$) \\ \hline 
\multicolumn{6}{c}{\textbf{Mean}} \\
 0 & 1.00 & 8.87e-4 & 4.86e-5 & 5.6e-3 \\
 1 & 1.00 & 1.05e-3 & 5.29e-5 & 5.9e-3  \\
 2 & 1.00 & 7.64e-4 & 5.49e-5 & 5.9e-3  \\
\multicolumn{6}{c}{\textbf{Standard Deviation}} \\
 0 & 1.31e-2 & 0.91e-4 & 5.61e-8 & 1.81e-4  \\
 1 & 1.48e-2 & 1.13e-4 & 5.67e-8 & 1.90e-4   \\
 2 & 1.20e-2 & 0.73e-4 & 3.13e-8 & 1.37e-4   \\
\multicolumn{6}{c}{\textbf{Pearson Correlation factor}} \\
 (0,1) & -0.49 & 1.22 & 1.47e-4 & 9.84e-3  \\
 (1,2) & -0.39 & 1.14 & 8.96e-5 & 7.77e-3  \\
 (2,0) & 0.03 & 0.98 & 2.62e-4 & 1.29e-2  \\
\end{tabular}
\end{ruledtabular}
\end{table*}

During training, the loss calculation, referenced in Equation \ref{Dkl gradient}, requires sampling from the target posterior distribution \( p(\beta | \mathbf{X}_d) \). This is achieved by applying Poisson fluctuations to each dataset \( \mathbf{X}_d \). Subsequently, for each varied dataset, we identify the most likely \( \beta \). The resulting \( \beta \) values from this process for a given dataset effectively sample the target posterior distribution. While this simulation step might be resource-heavy for latent spaces with high dimensions, it is a one-time requirement before training, conducted alongside the creation of the dataset. Consequently, one must balance the trade-off between the lengths of pre-training and training against the speed of sampling, which ultimately hinges on the particular scenario at hand. 

\section{Statistical and systematical errors in the statistical moment predictions}
\label{Statistical and systematical errors in the statistical moment predictions}

We estimate these errors by calculating the Mean Squared Error (MSE) between a statistic from the predicted posterior distribution \( S_\text{NF} \) and that of the target posterior distribution \( S_{\text{target}} \) for a given Asimov datapoint. We generate many samples of the two statistics using bootstrapping on the posterior datasets. In order to decouple the fluctuations of \( S_\text{NF} \) from the ones of \( S_\text{target} \), we compare \( S_\text{NF} \) to  \( \overline{S_\text{target}} \)  the average of the target statistic obtained with bootstrapping for a given Asimov datapoint. The MSE calculation is as follows: 

\begin{align}
\text{MSE} &= \mathbb{E}_\text{Boot}\left(\left|S_\text{NF} - \overline{S_\text{target}}\right|^2\right) \notag\\
&= V^\text{S}_\text{stat} + {V_\text{bias}^S}
\end{align}

 \( \mathbb{E}_\text{Boot} \) denotes the expectation over the bootstrapped posterior datasets.   The first term corresponds to the statistical error $V^\text{S}_\text{stat}(N)$ which typically scales with $\frac{1}{N}$, where $N$ is the number of posterior samples while the second term is constant as a function of $N$. Therefore, the MSE can be modeled by the function:
$$ \text{MSE}_\text{fit} (N) = \frac{a}{N} + b, \quad a,b>0$$

In order to estimate these fit parameters $a$ and $b$, we  calculate the MSE for increasing number of posterior samples and fit the curve with the above equation for each Asimov datapoint.  This is done using $200$ datasets generated from Asimov data points uniformly sampled from the $[0.5,1.5]^3$ volume. The errors for the studied statistics are summarized in Table \ref{tab:my-table}.

Our analysis shows that the error in the mean prediction is mainly systematical and tends also to be bias-dominated for the standard deviation for $N>>2,000$. More posterior samples are required for the error in the Pearson correlation factors to be dominated by the bias, between $4,000$ and $13,000$ posterior samples depending on the correlation factor.  The skewness and kurtosis measurements exhibit higher statistical error.  While the MSE presented in the table can be linked to intrinsical error of the model, the skewness and kurtosis values are not significant at a $0.05$ level, meaning that both \( p_\text{target}\) and \( q_\text{NF}\) are very close to Gaussian distributions.

\bibliography{apssamp}

\end{document}